\documentstyle[aps,prl,twocolumn,epsf,epsfig]{revtex}
\begin{document}

\draft
\wideabs{
\title{The semiclassical limit of chaotic eigenfunctions} 

\author{ Eduardo G. Vergini }

\address{ Departamento de F\'{\i}sica, Comisi\'on Nacional de
Energ\'{\i}a At\'omica. 
 Av. del Libertador 8250, 1429 Buenos Aires, Argentina.}


\date{\today}

\maketitle

\begin{abstract}
A generic chaotic eigenfunction has
a non-universal contribution consisting of scars of short
periodic orbits. This contribution, which can not be explained
in terms of random universal waves, survives the semiclassical limit
(when $\hbar$ goes to zero). In this limit, the sum of 
scarred intensities is a simple function of 
$\eta\equiv \!\sqrt{\pi /2}\; (f\!-\!1) h^{-1}_T (\sum \lambda_i^2)^{1/2} $,
with $f$ the degrees of freedom, $h_T$ the topological entropy and
$\{\lambda_i\}$ the set of positive Lyapunov exponents.
Moreover, the fluctuations of this representation go to zero as 
$1/|\ln \hbar|$. For this reasson, 
we will be able to provide a detailed description of a generic 
chaotic eigenfunction in the semiclassical limit.

\end{abstract}
}
\narrowtext



Berry \cite{Be1} and Voros \cite{Vo1} proposed a 
semiclassical description of chaotic eigenfunctions by
considering the surface of constant energy as the unique
classical invariant able to support them. This would be the
case if the time required for the definition of individual
eigenfunctions was infinite. However, the required Heisemberg
time $T_H$ is finite for finite values of $\hbar$, even though
it goes to infinity in the semiclassical limit (when $\hbar$
goes to zero).

Recently, we have derived a semiclassical theory which
describes individual eigenfunctions in terms of short periodic
orbits $(\bf POs)$ \cite{Ve1,Ve2,Ve3}. The number of $\bf POs$ is 
such that the sum of their periods is around $T_H$. 
Using these invariants, we will give a  
description of chaotic eigenfunctions with fluctuations 
going to zero with $\hbar$.

It is worth to emphasize that the fluctuations of the
Berry-Voros description go to infinity in the semiclassical
limit ({\bf SL}). But this statement demands explanation because it is
common to find in the literature expresions like {\it "
...this description is supported by Shnirelman theorem."}.
For example, the peaks of the Husimi distribution of a chaotic 
eigenfunction
have very strong  fluctuations with respect to their classical
ergodic measure; by assuming a random wave hypothesis they 
go logarithmically to infinity as $\hbar \rightarrow 0$ \cite{No1}.
However, each of the peaks occupy a volume 
(the semiclassical volume of a quantum state)
going to zero with $\hbar$. Then, a smoothing on
a region independent of $\hbar$, a classical smoothing, is
sufficient for washing out fluctuations in the
{\bf SL}. In this context,
Shinirelman theorem \cite{Sh1} may be expresed as
follows: {\it a classical smoothing is sufficient for the 
elimination of fluctuations (with respect to the classical
ergodic measure) of generic chaotic eigenfunctions 
in the {\bf SL}.}

We are going to show that the highest peaks of the Husimi are not 
completely random. On the contrary, they live (within an error of 
order $\sqrt{\hbar}$ in each
direction) on short {\bf POs}, and
their phases (considering in this case
the corresponding Bargmann function \cite{No1}) are connected 
semiclassically. 
That is, high fluctuations correspond with the scars of short {\bf POs} 
introduced by Heller \cite{He1} 
(although the meaning of {\it short} is other; see Eq. (\ref{crit})).
In this respect, the following commonly used expression is definitely
wrong: {\it "...scarred eigenfunctions do not contradict 
Shnirelman theorem because they are of null measure."}. Actually, 
scar phenomena  are of a semiclassical nature (they manifest on scales going
to zero with $\hbar$). Then again, a classical smoothing washes out
these very rich structures in the {\bf SL}. 

Bogomolny \cite{Bog} and Berry \cite{Be2} incorporated the idea of
scars into the description of chaotic eigenfunctions.
These authors concluded that chaotic eigenfunctions consist of a
dominant universal contribution, decorated by scars of short {\bf POs};
but this result was obtained after making an average over an energy 
interval that also eliminates fluctuations.
In our study, we consider individual eigenfunctions, 
without either a classical smoothing nor an energy averaging. Then,
we arrive to the conclusion that a generic chaotic eigenfunction is
scarred by a set of short {\bf POs} which characterizes the state, and
the sum of their intensities survives the {\bf SL}. This fact stresses
that localization on short {\bf POs} is the signature of chaos
in this limit. Moreover, we will give the mean value and dispersion
of the scarred intensities showing that relative fluctuations go to 
zero as $1/| \ln{\hbar} |$, and providing a detailed description
of a generic chaotic eigenfunction.


We have developed a systematic semiclassical construction of wave functions
living in the neighbourhood of unstable {\bf POs}. Reference 
\cite{Ve1} provides the construction of resonances without transverse
excitations and ref. \cite{Ve2} applies the recipe to the Bunimovich stadium
billiard. In ref. \cite{Ve3} we construct resonances of a {\bf PO}
with transverse excitations, all at the same Bohr-Sommerfeld ({\bf BS})
quantized energy. Then,
scar functions are the result of a minimization of the energy dispersion in 
this basis. Finally, ref. \cite{Ca1} gives the details for the construction
of scar functions in the stadium billiard.
Scar functions are the objects on which we are going to focus our
description. We are thinking about structures living along the 
manifolds up to the first homoclinic point. Then,
it is not necessary to include interference effects, and the description
remains essentially simple. 


Let $\gamma$ be an unstable $\bf PO$ of the system with period $T_{\gamma}$
and Lyapunov exponent $\lambda_{\gamma}$ per unit time (consider
for the moment a conservative Hamiltonian system with two degrees of
freedom). Let $\phi_{\gamma}$
be the corresponding scar function with {\bf BS} energy $E_{\gamma}$.
Finally, let $\varphi_{\mu}$ be the set of normalized eigenfunctions of
the system with eigenenergies $E_{\mu}$. In ref. \cite{Ve3} we have
shown that the set of intensities 
$I_{\mu}=|\langle \phi_{\gamma} |\varphi_{\mu} \rangle |^2$ 
semiclassically satisfies, $\sum_{\mu} I_{\mu} =1 $, 
$\sum_{\mu} E_{\mu} I_{\mu} = E_{\gamma}$, and 
\begin{equation}
 \sigma_{\gamma} \equiv \sqrt{{\sum}_{\mu} (E_{\mu}-E_{\gamma})^2 I_{\mu}}
 = \hbar \lambda_{\gamma} \Gamma /2 \cite{nota}.
\label{sigma}
\end{equation}
In the {\bf SL}, the universal dispersion $\Gamma$  goes
to zero as  $2 \pi / |\ln \hbar | $
(expresions of $\Gamma$ for finite values of $\hbar$ are given in 
ref. \cite{Ve3}).
Then, the life time $\tau _{\gamma}\equiv \hbar/\sigma_{\gamma}$ of
$\phi_{\gamma}$ diverges logarithmically in the same way as the
Ehrenfest time does (but they are different times). This extremely low
decay \cite{decay} is governed by a Gaussian law 
($|\langle \phi_{\gamma}(0)|\phi_{\gamma}(t)\rangle|^2=
e^{-t^2/\tau_{\gamma}^2}$), and 
the smooth part of the intensities $I_{\mu}$ (the strength function)
results a Gaussian function. Then, defining
$\;\epsilon \equiv (E-E_{\gamma}) \rho_E$ (with $\rho_E$ the energy density),
the mean value of the intensities at $\epsilon$ is 
\begin{equation}
\overline{I}(\epsilon)=e^{-\epsilon^2 /2 n^2}/\sqrt{2 \pi n^2},
\label{mean}
\end{equation}
where $n\equiv \sigma_{\gamma} \rho_E$ is the mean number of eigenenergies
contained in one energy dispersion.

Another consequence of the Gaussian decay is that Eq. (\ref{sigma})
is also valid for systems with $f$ degrees of freedom if we replace
$\lambda_{\gamma}$ by $(\sum \lambda_i^2)_{\gamma}^{1/2}$, with 
$\{ \lambda_i \} _{\gamma}$ the set of $f-1$ positive Lyapunov exponents
of $\gamma$ (for an exponential decay the change would be
$\lambda_{\gamma}$ by the K-S entropy $(\sum \lambda_i)_{\gamma}$).

Now, we are able to establish a criterium in order to decide when a
$\bf P.O$ is short. We propose a relation of the form 
$T_{\gamma} < \beta \tau_{\gamma}$, with $\beta$ a constant to be
determined from the theory of short {\bf POs}. This theory says
that in the {\bf SL} an eigenfunction is defined by {\bf POs}
of period lower than $T_0= h^{-1}_T \ln ( T_H h_T )$ \cite{Ve1},
with $h_T$ the topological entropy. Then, in the {\bf SL}
$T_0$ would be equal to $\beta$ times $\tau$
(the life time of a generic scar function), with
\begin{equation}
\tau=\pi^{-1}  (\sum \lambda_i^2)_{sys}^{-1/2}\;| \ln \hbar |.
\label{tau}
\end{equation}
$\{ \lambda_i \} _{sys}$ is the set of positive Lyapunov exponents
of the system. Finally, we say that a {\bf PO} $\gamma$ is short when
\begin{equation}
T_{\gamma}/\tau_{\gamma}< \sqrt{2 \pi}\; \eta \;\;\;"short\; {\bf PO}
\;condition",
\label{crit}
\end{equation}
where the classical invariant
\begin{equation}
\eta\equiv \sqrt{\pi /2} (f\!-\!1)h^{-1}_T (\sum \lambda_i^2)_{sys}^{1/2},
\label{eta}
\end{equation}
will play a central role below. Evidently, any {\bf PO} is
short at sufficiently high energies. 

Of all the intensities of a given scar function we are 
actually interested in those with the highest values (they are 
certainly related to the phenomenon of localization on short {\bf POs}).
We are going to study the properties of
these  high intensities in terms of a statistical model. The main
purpose is to estimate their mean values and dispersions, and to
provide a range in the spectrum where they live.

The fluctuations of an intensity $I$ at $\epsilon$, around its mean
value $\overline{I}(\epsilon)$ (see Eq. (\ref{mean})), will be 
described as usual \cite{porter}
by a chi-squared distribution with one degree of freedom 
for systems with time reversal symmetry (appendix D treats systems
without time reversal).
Then, the probability of finding an intensity lower than $I$ (the
accumulated probability) is given, for 
$a\equiv I/ \overline{I}(\epsilon)>1$, by
$\;F_{\epsilon}(I) = 1- \sqrt{ 2 /\pi a}\; e^{-a/2}\;
(1-a^{-1}+3 a^{-2} + \cdots )\;$.
In order to obtain the distribution for an arbitrary intensity 
(independent of its position in the spectrum), 
we make an averaging; if $N$ is the number of
intensities contributing to the scar function,  
$\;F(I)\equiv (1/N) \int_{-N/2}^{N/2}  F_{\epsilon}(I) d\epsilon$.
This expresion is computed in terms of Gaussian integrals after
the following expansion:
$e^{-a/2}=e^{-y} e^{-y\epsilon^2/2n^2} (1-y\epsilon^4/8n^4 +\cdots)$,
where $y\equiv \sqrt{\pi/2}\; n I$ (note that 
$\epsilon^2/n^2={\cal O}(y^{-1})$). Then, 
\begin{equation}
F(I)=1-\frac{\sqrt{2} n}{N} \frac{e^{-y}}{y} \left( 1- \frac{9}{8 y}
+ \frac{305}{128 y^2} + \cdots \right).
\label{densidad}
\end{equation}

Now, if $x_1$ is the greatest of the intensities, $x_2$ the second one
and so on, the probability density of $x_j$ is given by Eq. (\ref{prob}).
A simple estimation of $x_j$ is derived from
$1-F(x_j)=j/N$. With this in mind, we define the random variable $z_j$ 
by the relation
\begin{equation}
1-F(x_j)=e^{-z_j} (j/N).
\label{defz}
\end{equation}
Combining Eqs. (\ref{densidad}) and (\ref{defz}), $x_j$ is given by
\begin{equation}
y_j \equiv \sqrt{\pi /2} n x_j\simeq \alpha-\ln (\alpha+9/8) + b +b^2/2,
\label{ymedio}
\end{equation}
with $\alpha \equiv z_j+\ln(\sqrt{2}n/j)$ 
and $b\equiv\ln (\alpha+287/128)/(\alpha+17/8)$. 
Equation (\ref{ymedio}) works for $\alpha >1$.

Equation (\ref{med}) gives the mean value $\overline{z_j}$ and 
dispersion $\sigma_{z_j}$ of
$z_j$. With these, the mean value of $y_j$ is obtained
from Eq. (\ref{ymedio}) by setting $\overline{\alpha}$ 
($=\overline{z_j}+\ln(\sqrt{2}n/j)$)
in place of $\alpha$ and adding to the rhs the term
$c\equiv\sigma_{z_j}^{2}/2 (\overline{\alpha}+9/8)^2$ (because 
$\overline{\ln(\alpha+9/8)}\simeq\ln(\overline{\alpha}+9/8)-c$). On 
the other hand, the dispersions of $y_j$ and $z_j$ are
equal to the leading order. Then, we arrive to the first
conclusion: {\it relative fluctuations go to zero in the {\bf SL}
as follows},
\begin{equation}
\sigma_{x_j}/x_j \sim 1/[\sqrt{j} 
\ln(\sqrt{2}n/j)]={\cal O}(1/| \ln \hbar |).
\label{relative}
\end{equation}

The next question is to know where $x_j$ can be found. The probability
$p(\epsilon)$ of finding $x_j$ near $\epsilon$ is proportional to
the probability density $dF_{\epsilon}/dI$ at $x_j$; that is,
$\;p(\epsilon) \propto
e^{-(y_j+1/2)\epsilon^2/2n^2} (1-y_j \epsilon^4/n^4+ \dots)\;$.
Then, $x_j$ is restricted to a range 
$\Delta E_j\simeq \sigma_{\gamma}/\sqrt{y_j+1/2}$ around  $E_{\gamma}$
which in units of $2\pi\hbar/T_{\gamma}$ (the distance between 
consecutive {\bf BS} quantized energies) goes to zero as 
$| \ln \hbar |^{-3/2}$. In conclusion: {\it scars of $\gamma$ accumulate 
in the vicinity of $E_{\gamma}$} in the {\bf SL}. 

In the following, we are going to discuse a fundamental question. 
How many universal waves participate in the description of a chaotic
eigenfunction?. For maps the question is simple; there are
$N_u$ waves with $N_u$ equal to the area of the map divided by
$2 \pi \hbar$. However, for Hamiltonian systems it is not clear
in general which one is the right, if any, Poincar\'e surface
of section.

At classical level, we can say that during an evolution for a time
lower than the period $T_{min}$ of the shortest {\bf PO}, it is
impossible to extract non-universal information of the system. Using this
time, Berry \cite{Be3} distinguishes 
universal aspects of the spectral rigidity
from the non-universal ones.
However, $T_{min}$ is not very useful because it is
simply an estimation.
Evidently, we need a time
representing a short universal evolution but such that its determination
is based on a deep knowledge of the system. For these reasons we propose
the inverse of the topological entropy for this time. Then, if an 
eigenfunction is defined after an evolution equal to $T_H$ and we
can do the travel in steps no greater than $h_T^{-1}$, where each
step defines a universal wave, the number of required universal waves is
\begin{equation}
N_u = T_H h_T.
\label{universal}
\end{equation}
There is a way of verifying the accuracy of Eq. (\ref{universal}).
In billiards, Birkhoff coordinates define the right Poincar\'e surface
of section (because all classical or quantal information is 
contained on the boundary). The area of the section is 
$2\hbar k {\cal L}$, with
${\cal L}$ the lenght of the boundary and $k$ the eigenwave number.
Then, $N_u=k {\cal L}/ \pi$, and assuming that also
Eq. (\ref{universal}) is right, we arrive to the following expresion
for the topological entropy $h'_T$ per bounce
\begin{equation}
h'_T={\cal L} \bar{l} /\pi {\cal A},
\label{topological}
\end{equation}
with ${\cal A}$ the area of the billiard and $\bar{l}$ the
mean length per bounce. We have verified Eq. 
(\ref{topological}) within an error of $2\%$ 
in the stadium billiard with radius unity and several different areas.

So far, we have described scar functions in the basis of eigenfunctions.
Now according to the theory of short {\bf POs} \cite{Ve1}, eigenfunctions
would be described equivalently in terms of scar functions. That is,
imagine the spectrum of scar functions given by all {\bf BS} energies
of all short {\bf POs}. This means for a given {\bf PO}, all {\bf BS}
energies in those energy regions where $\gamma$ is short (following the
criterium of Eq. (\ref{crit})). Then, an eigenfunction $\varphi_{\mu}$
with eigenenergy $E_{\mu}$ is represented in the basis of scar functions
by the intensities
$I_{\gamma}=|\langle \varphi_{\mu} | \phi_{\gamma}\rangle |^2$,
which are concentrated around $E_{\mu}$ in the
spectrum of {\bf BS} energies. The dispersion
$\sigma_{\mu} \equiv (\sum (E_{\gamma}-E_{\mu})^2 I_{\gamma})^{1/2}$
depends on the Lyapunov exponents of the orbits with {\bf BS} energies
in the vicinity of $E_{\mu}$. However, in the {\bf SL} there is a 
uniformization and the energy dispersion results independent of the
position in the spectrum and equal to $\hbar/\tau$, with $\tau$ given
by Eq. (\ref{tau}).
The smooth part of the intensities $I_{\gamma}$
is given by Eq. (\ref{mean}), with 
\begin{equation}
n=\sigma \rho_E=T_H/2\pi \tau,
\label{ene}
\end{equation}
and $\epsilon=(E-E_{\mu}) \rho_E$ \cite{nota2}.
If $x_1$ is the highest of the intensities, $x_2$ the second one and
so on, the mean value of $x_j$ is given by Eq. (\ref{ymedio}) (and
discusion thereafter), and the relative dispersion by Eq. (\ref{relative}).
Of all short {\bf POs}, those with possibility of having intensity $x_j$
satisfy $|E_{\mu}-E_{\gamma}|< \sigma/\sqrt{y_j+1/2}$.

The following question is to decide a criterium for scarring.
We will say that $\varphi_{\mu}$ is scarred by $\gamma$ if 
$I_{\gamma}$ is greater than the greatest of the intensities provided by
a random model of universal waves. The highest universal intensity is,
to the leading order, $I^{(u)}= 2 \ln(N_u)/N_u $ \cite{nota3}.
Then, using Eqs. (\ref{eta}), (\ref{universal}) and (\ref{ene}), the 
condition $x_j > I^{(u)}$ reduces to
\begin{equation}
{y}_j > \eta\;\;\; "scarring\; condition".
\label{condition}
\end{equation}
The number $n_{scar}$ of {\bf POs} satisfying Eq. (\ref{condition}) is
given by
\begin{equation}
\frac{n_{scar}}{n} \simeq e^{-\eta} \sqrt{\frac{8 \ln (1+1/\delta \eta)}
{(1+4\eta/ \delta)}},
\label{scar}
\end{equation}
with $\delta=(9+\sqrt{73})/2$. This formula (and the next too) interpolates 
the behaviors for
large $\eta$ (obtained from Eqs. (\ref{densidad}) and (\ref{defz}))
and $\eta$ going to zero (see appendix C).
 Finally, the sum of scarred intensities results
\begin{equation}
\sum_{j=1}^{n_{scar}} x_j \simeq \frac{2}{\sqrt{\pi}}\; e^{-\eta}-
\left( \frac{2}{\sqrt{\pi}} - 1 \right) e^{-2 \eta}.
\label{suma}
\end{equation} 

We emphasize that the unexpected results of Eqs. (\ref{scar}) and (\ref{suma})
depend decisevely on the use of scar functions in order to measure localization
on short {\bf POs}. On the contrary, by using wave packets in the transverse
direction to the motion (the so called vacuum states in ref. \cite{Ve3}),
there result $n_{scar}/n=0$ and $\sum x_j=0$ in the {\bf SL}. In this respect,
the departure of universal behavior found in ref. \cite{Kap}
has a weight going to zero in the {\bf SL}.

We stress that all formulae of statistical nature
verify an impresive agreement with numerical simulations. Moreover,
Eqs. (\ref{sigma}) and (\ref{mean}) were 
extensively verified in the stadium billiard. Finally, we present
an example where all the ideas developed in the article have been tested.
Figure 1) shows the decomposition of a very high excited chaotic state
(plotted in configuration space in ref. \cite{Ve3}). 
We found $9$ scarred short {\bf POs} ($n_{scar}\simeq 8.5$ from Eq. (\ref{scar}))
providing a contribution of the $38 \%$ (Eq. (\ref{suma}) predicts $33 \%$).
Moreover, scarred intensities are in good agreement with predictions (see
Fig. 1d)).

In conclusion, localization on short {\bf POs} survives the {\bf SL}
and depends exclusively on $\eta$. Evidently, this localization will
be strong in systems with few degrees of freedom. In particular for 
$f=2$, and assuming that $\lambda\simeq h_T$, this localization results
a universal property. 

This work was partially supported by SETCYP-ECOS A98E03.

Appendix A.
Let $I_1, I_2, \ldots , I_N$ be a set of independent random variables,
with common probability density $f(I)$, living in the range $[0,\infty)$.
Let $x_1$ be the greatest of the variables, $x_2$ the second one an so on.
The joint probability density is given by
$\;p(x_1, x_2, \ldots , x_N)=N! f(x_1) f(x_2)  \ldots  f(x_N)\;$,
for $x_1\geq x_2 \geq \ldots \geq x_N$; otherwise $p=0$. 
Then, the probability density of $x_j$ results (after integration over 
the others variables)
\begin{equation}
p (x_j )= \frac{N! f(x_j)\; F(x_j)^{N-j}\;
[1-F(x_j)]^{j-1}}{(N-j)! (j-1)!},
\label{prob}
\end{equation}
with $F(x)=\int_0^{x} f(x') dx'$ the accumulated probability.

Appendix B. We will derive the mean value and dispersion 
of the random variable $z_j$ defined in Eq. (\ref{defz}).
Assuming that $z_j={\cal O}(1)$, there results from Eq. (\ref{defz}) 
$1-F(x_j) ={\cal O} (j/N)$, and then
$F(x_j)^{N-j}=\exp{[(N-j)\ln F(x_j)]}=\exp{[-N (1-F(x_j))+{\cal O}(j^2/N)]}$.
Moreover, $N!/(N-j)!=N^j [1+{\cal O}(j^2/N)]$. Using these approximations
in Eq. (\ref{prob}), the probability density of $z_j$ is given by  
\[
p(z_j) \simeq j^j  e^{-j (z_j+e^{-z_j})}/(j-1)!.
\]
For instance, a numerical computation gives $\overline{z_1}\simeq 0.577$
and $\sigma_{z_1} \simeq 1.28$. On the other hand, for large values of
$j$ we change to a new variable $w=\sqrt{j} z_j$, and expanding in
powers of $1/\sqrt{j}$ there results
\[
p(w)\simeq \frac{e^{-w^2/2}}{\sqrt{2\pi}} \left( 1-\frac{1}{12j} \right)  
\left( 1+\frac{a}{\sqrt{j}}+\frac{b}{j}+\frac{c}{j^{3/2}}\right),
\]
with $a=w^3/6$, $b=w^4(w^2-3)/72$ and $c=w^5(5w^4-45w^2+54)/6480$.
Finally, using this density we have
\begin{equation}
\overline{z_j}\simeq \frac{1}{2j} + \frac{1}{12j^2} 
\;\;\;\;{\rm and}\;\;\;\;\sigma_{z_j}\simeq \frac{1}{\sqrt{j}} 
\left( 1+ \frac{1}{4j} \right),
\label{med}
\end{equation}
in excellent agreement with numerical data.

Appendix C. For $\eta$ going to zero, $n_{scar}$ satisfies $\overline{I}(n_{scar}/2)=I^{(u)}=\sqrt{2/\pi} \eta/n$. Then, using
Eq. (\ref{mean}) there results $n_{scar}/n \simeq 2^{3/2} \sqrt{\ln(1/\eta)}$.

Appendix D. The corresponding formulae for systems without time reversal
are the following:
$y_j \equiv \sqrt{2\pi} n x_j\simeq \alpha-\ln (\alpha+3/4)/2 + b +b^2$,
in place of Eq. (\ref{ymedio}), with 
$\alpha\equiv z_j + \ln(\sqrt{2 \pi} n/j)$ and 
$b=\ln(\alpha+31/16)/(4\alpha+5)$.
Equation (\ref{condition}) is the same. Equation (\ref{scar}) is replaced
by
\[
n_{scar}/n\simeq e^{-\eta}\; \sqrt{2 \pi \ln(1+1/\eta)},
\]
and Eq. (\ref{suma}) by
\[
\sum_{j=1}^{n_{scar}} x_j \simeq \frac{(1+\eta)}{\sqrt{3/4+\eta}}\; 
e^{-\eta}-\left( \frac{2}{\sqrt{3}} - 1 \right) e^{-2 \eta}.
\]


\begin{figure}[tbp]
\centering \leavevmode
\epsfxsize=4cm
\center{\epsfig{file=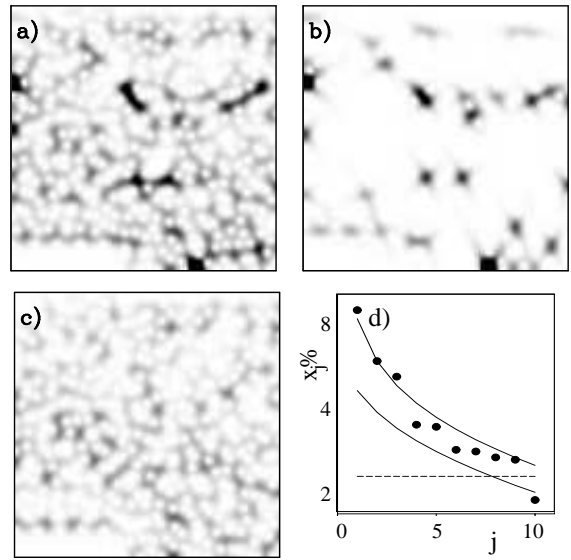, ,width=7.5cm,angle=0}}
\vspace{0.5cm}
\caption{Linear Husimi density plots of: a) the state number 141,755 of the
desymmetrized stadium billiard with radius $1$ and area $1+\pi/4$, b) its
non-universal contribution consisting of $9$ scar functions, and c)
its universal contribution consisting of $819$ plane waves. d) The set
of scarred intensities (dots) and the theoretical estimation curves
$\overline{x_j} \pm \sigma_{x_j}$; horizontal line 
displays the value of the highest universal intensity. }
\label{fig1}
\end{figure}


\begin{references}
\bibitem{nota} Equation (\ref{sigma}) is derived from Eq. (40) of ref. \cite{Ve3} by considering that the rate of divergence is uniform in time.
\bibitem{decay} With respect to the characteristic classical decay which
is governed by the Lyapunov exponent of the system.
\bibitem{nota3} This result is obtained by the method
developed in this article, setting $\overline{I}=1/N_u$ in place of
Eq. (\ref{mean}).
\end{references}
\end{document}